\documentclass[twocolumn,aps]{revtex4}

\pdfpageattr {/Group << /S /Transparency /I true /CS /DeviceRGB>>}

\usepackage{graphicx} 
\usepackage{overpic}

\newcommand{\ket}[1]{\left|#1\right\rangle}

\begin{document}

\title{Low-overhead surface code logical $H$}

\author{Austin G. Fowler}

\affiliation{Centre for Quantum Computation and Communication Technology,\\
School of Physics, The University of Melbourne, Victoria 3010, Australia}

\date{\today}

\begin{abstract}
We present an improved low-overhead implementation of surface code logical $H$. We describe in full detail logical $H$ applied to a single distance 7 double-defect logical qubit in an otherwise idle scalable array such qubits. Our goal is to provide a clear description of logical $H$ and to emphasize that the surface code possesses low-overhead implementations of the entire Clifford group.
\end{abstract}

\maketitle

It has been noted previously that the surface code permits low-overhead implementations of logical $H$ \cite{Fowl08} and logical $S$ \cite{Alif07,Jone10} implying it can efficiently implement the entire Clifford group as it possesses an efficient logical CNOT. Here we present an improved implementation of logical $H$ and describe it in far greater detail that has been done before. We assume basic familiarity with surface code quantum computing \cite{Raus07,Raus07d,Fowl08}. We have chosen an explicit presentation of a distance 7 example over a generic arbitrary distance presentation as we believe the distance 7 example is minimum size when assuming a lattice of equally spaced logical qubits. Lower distance logical $H$, including distance 3, is possible, however more space is required around the manipulated distance 3 logical qubit than is available in a minimum spacing lattice of distance three logical qubits.

\begin{figure*}
\begin{center}
\resizebox{150mm}{!}{\includegraphics{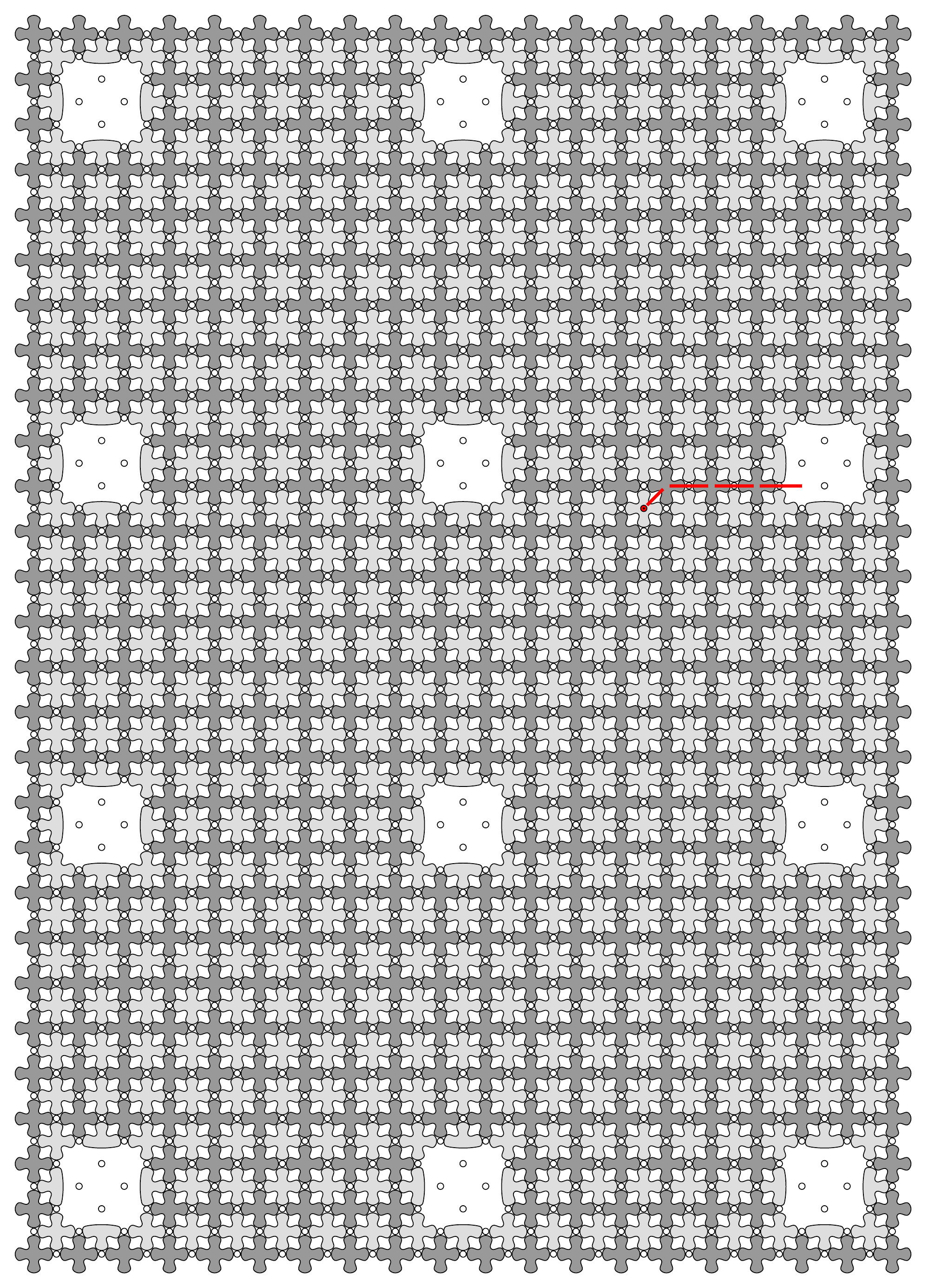}}
\end{center}
\caption{Initial configuration of dual (smooth) defects. The center two defects represent the logical qubit we will be applying logical $H$ to. White circles represent data qubits. Dark patches represent $Z$-stabilizers. Light patches represent $X$-stabilizers. The regular array of double-defect logical qubits has been sized so as to realize a distance 7 quantum error correction code. The surface is assumed to extend a significant distance in all directions. 3.5 segments of an error chain are shown in red. The last half segment runs diagonally into the future \cite{Wang11}.}\label{Hadamard_A}
\end{figure*}

\begin{figure*}
\begin{center}
\resizebox{150mm}{!}{\includegraphics{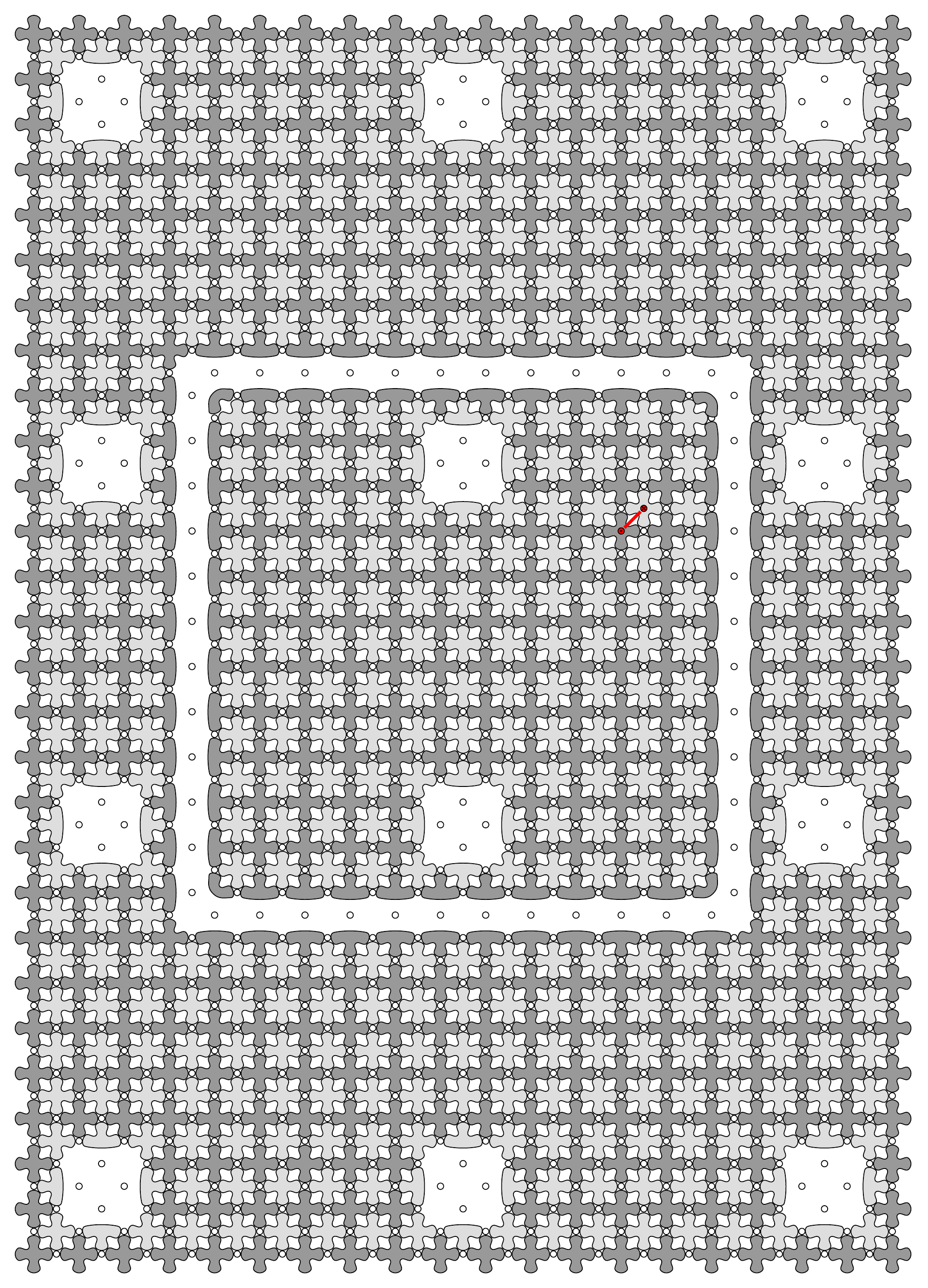}}
\end{center}
\caption{A number of $Z$-stabilizers have been reduced in weight to create a double ring of primal (rough) boundary. Note that rough boundaries can be as close as shown to the smooth defects without reducing the distance of the code as the minimum weight logical operators encircling the smooth defects remain weight 8. The data qubits between the two new rough boundaries are measured in the $Z$ basis. Half a segment of an error chain runs diagonally into the past. This measurement pattern is repeated twice more adding two more temporal segments to the error chain bringing the total number of segments to 6.}\label{Hadamard_B}
\end{figure*}

\begin{figure*}
\begin{center}
\resizebox{150mm}{!}{\includegraphics{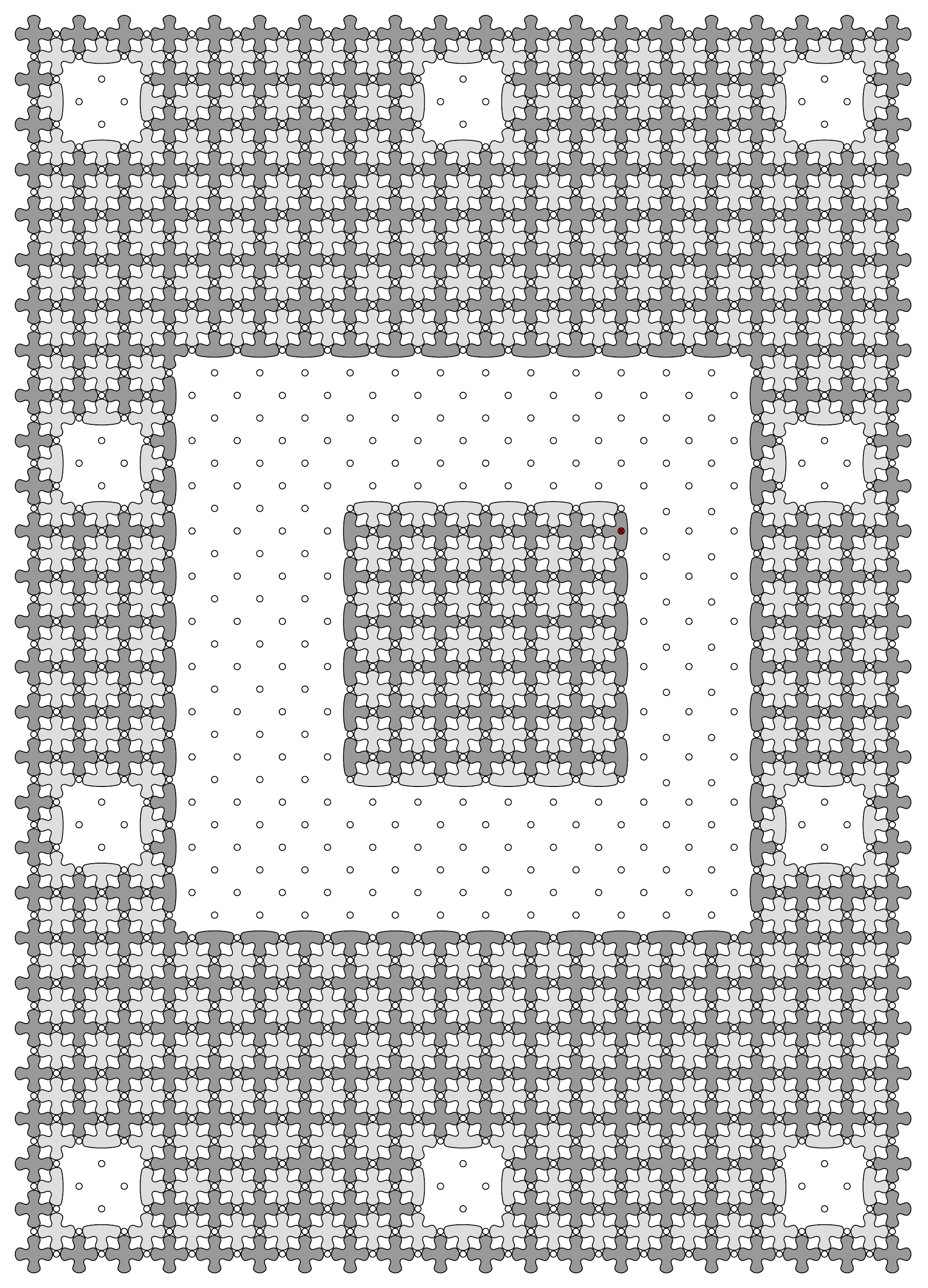}}
\end{center}
\caption{A large number of stabilizer checks are now turned off or reduced in weight to create a distant 7 square logical qubit. Logical data is preserved. The end of a seventh segment of the error chain is shown, demonstrating that we still have a distance 7 code as there are no shorter error chains than this. The data qubits adjacent to the upper and lower edge of the square are measured in the $X$ basis, all others are measured in the $Z$ basis.}\label{Hadamard_C}
\end{figure*}

\begin{figure*}
\begin{center}
\resizebox{150mm}{!}{\includegraphics{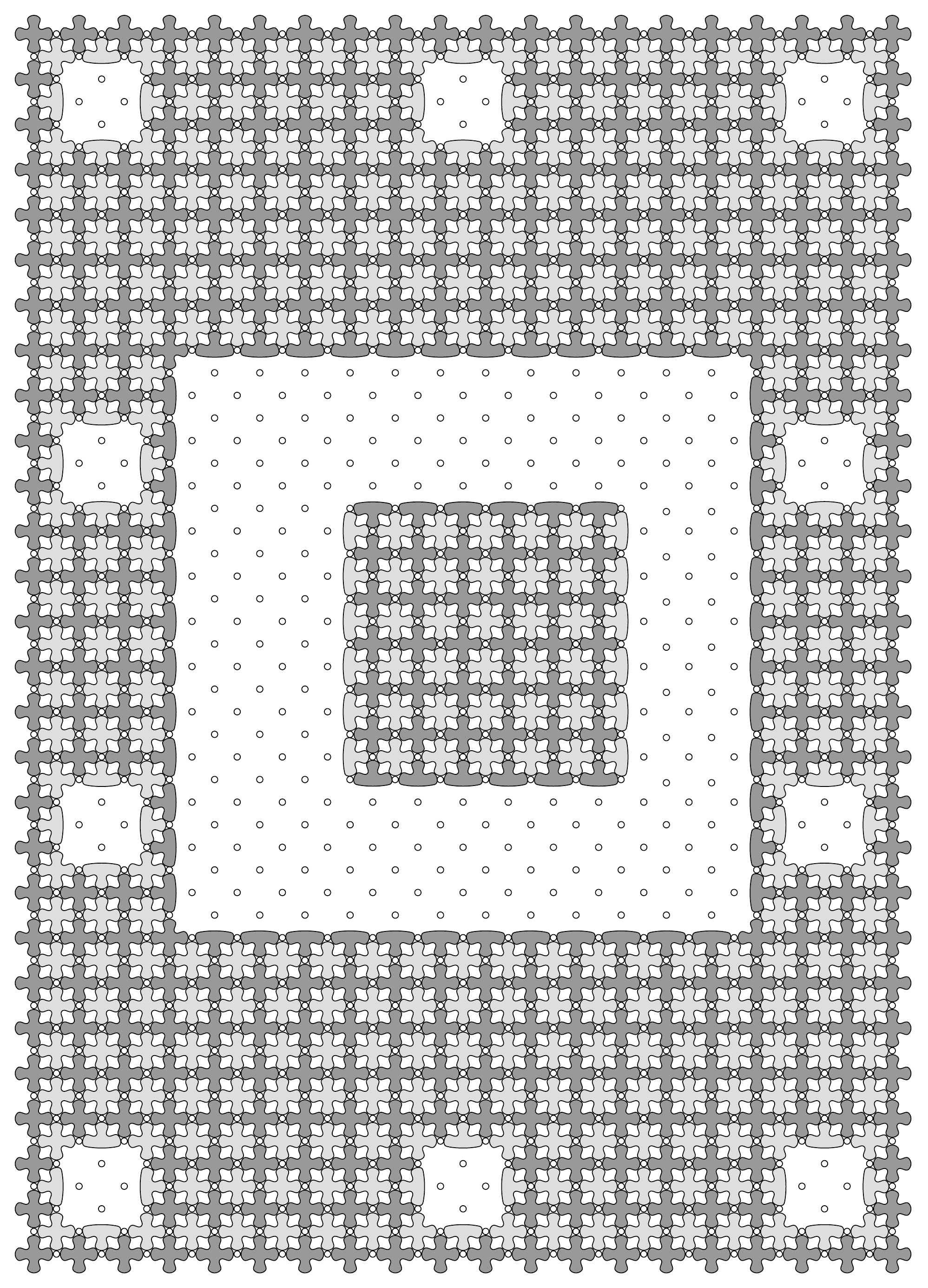}}
\end{center}
\caption{Transversal physical $H$ is now applied to each of the data qubits in the square logical qubit. This can be done during measurement and initialization of the syndrome qubits and hence takes no real time. In addition to implementing logical $H$, this converts $Z$-stabilizers into $X$-stabilizers and vice versa. If the surface remained in this configuration, in future rounds of error detection different circuits would be applied across the square. Note, however, that these error detection circuits are not applied in this step as the transversal $H$ takes no real time.}\label{Hadamard_D}
\end{figure*}

\begin{figure*}
\begin{center}
\resizebox{150mm}{!}{\includegraphics{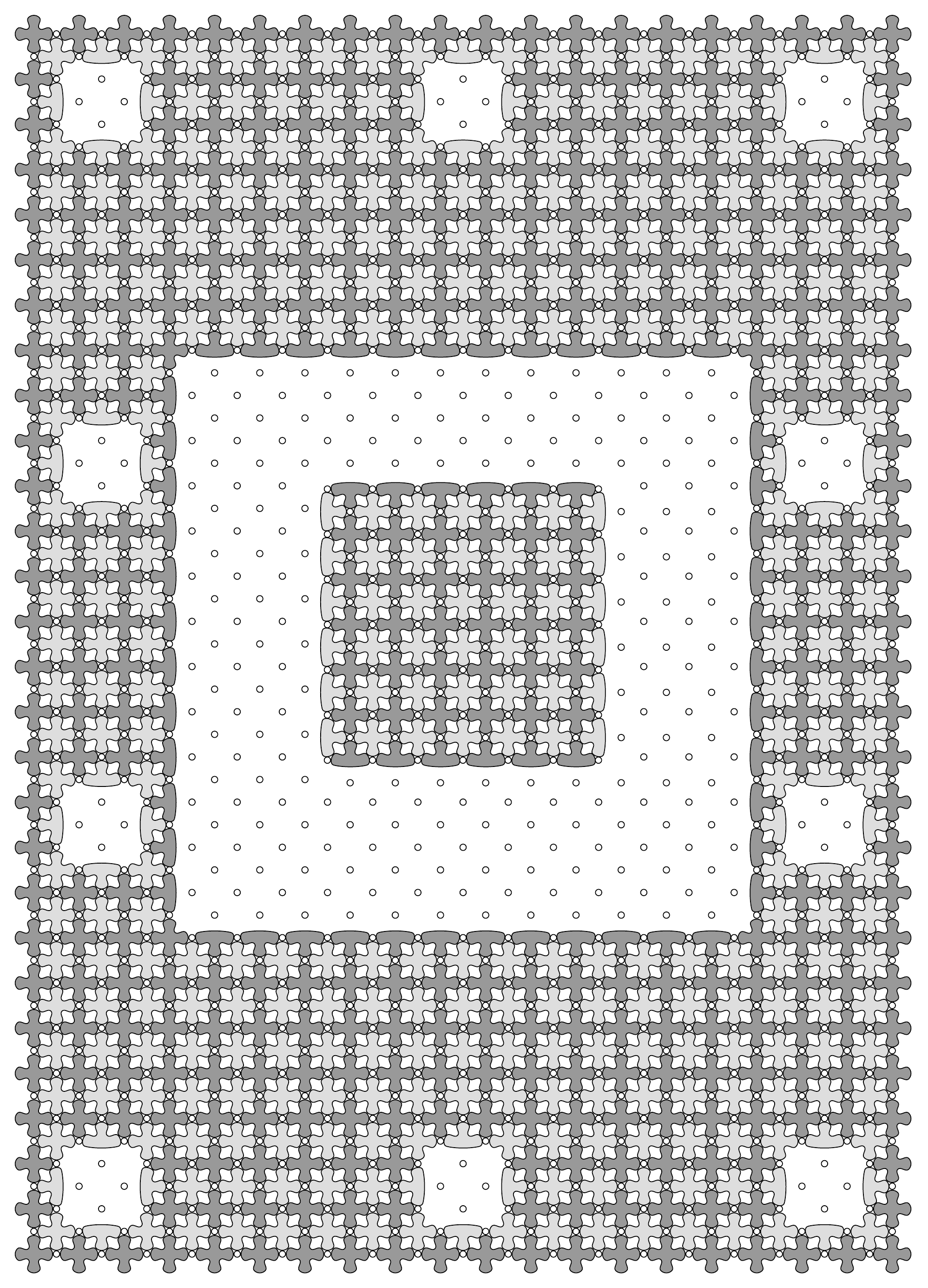}}
\end{center}
\caption{Swap gates are used to shift the square logical qubit data qubits up and left, realigning the stabilizers with the regular pattern of the rest of the computer. These swap gates do not introduce correlated errors between data qubits, however they would take sufficient time that while the surface outside the outer primal (rough) boundary would perform a round of error detection as normal, those in the interior square would do nothing other than these swap gates and identity gates to resynchronize. This would not change the threshold error rate of the scheme, as this missing round of error detection is a single localized event in time.}\label{Hadamard_E}
\end{figure*}

\begin{figure*}
\begin{center}
\resizebox{150mm}{!}{\includegraphics{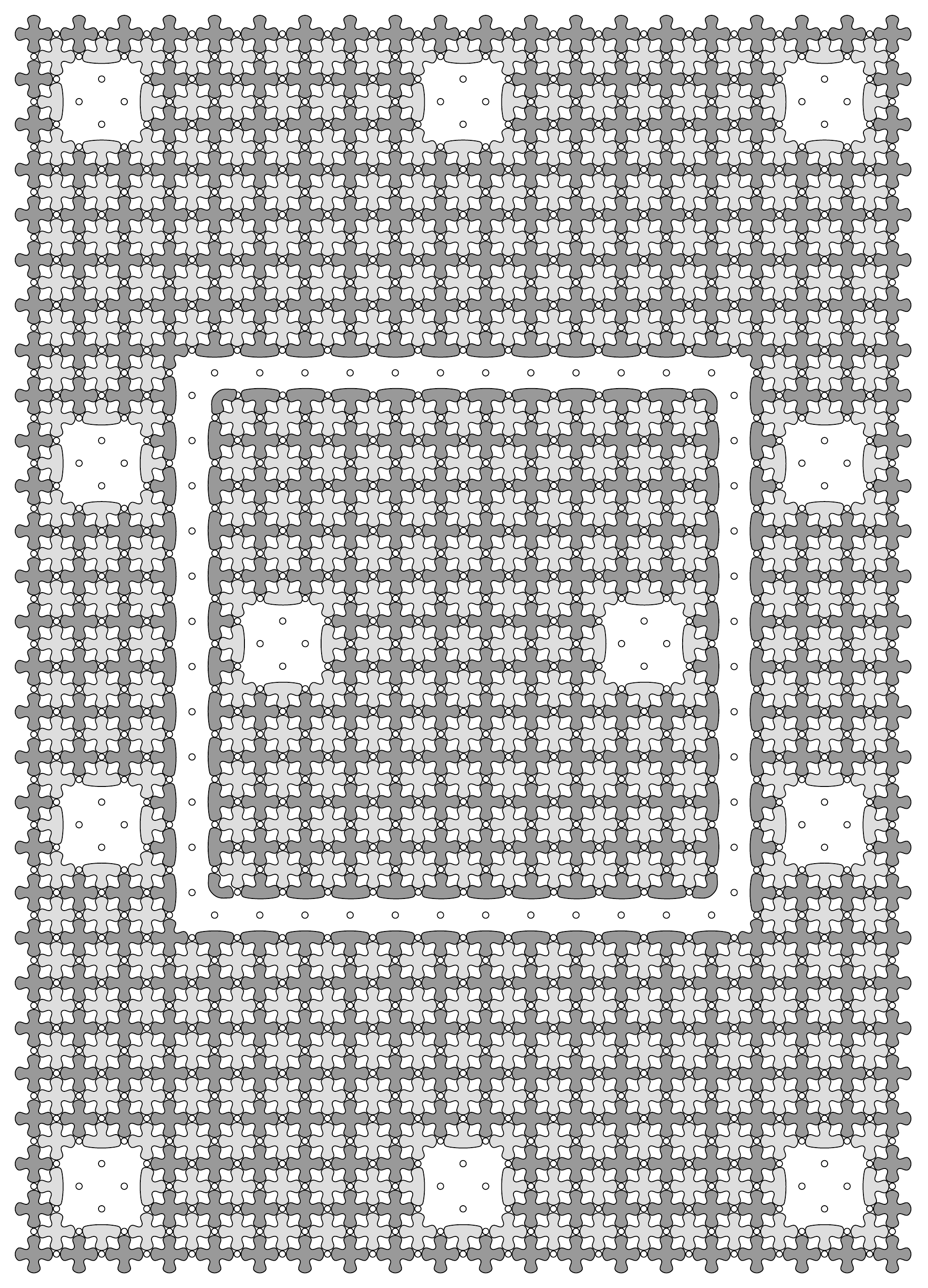}}
\end{center}
\caption{Many stabilizer checks are turned back on creating a new double smooth defect logical qubit preserving the logical information. All newly reintegrated data qubits are initialized to $\ket{0}$, effectively creating a temporal primal boundary.}\label{Hadamard_F}
\end{figure*}

\begin{figure*}
\begin{center}
\resizebox{150mm}{!}{\includegraphics{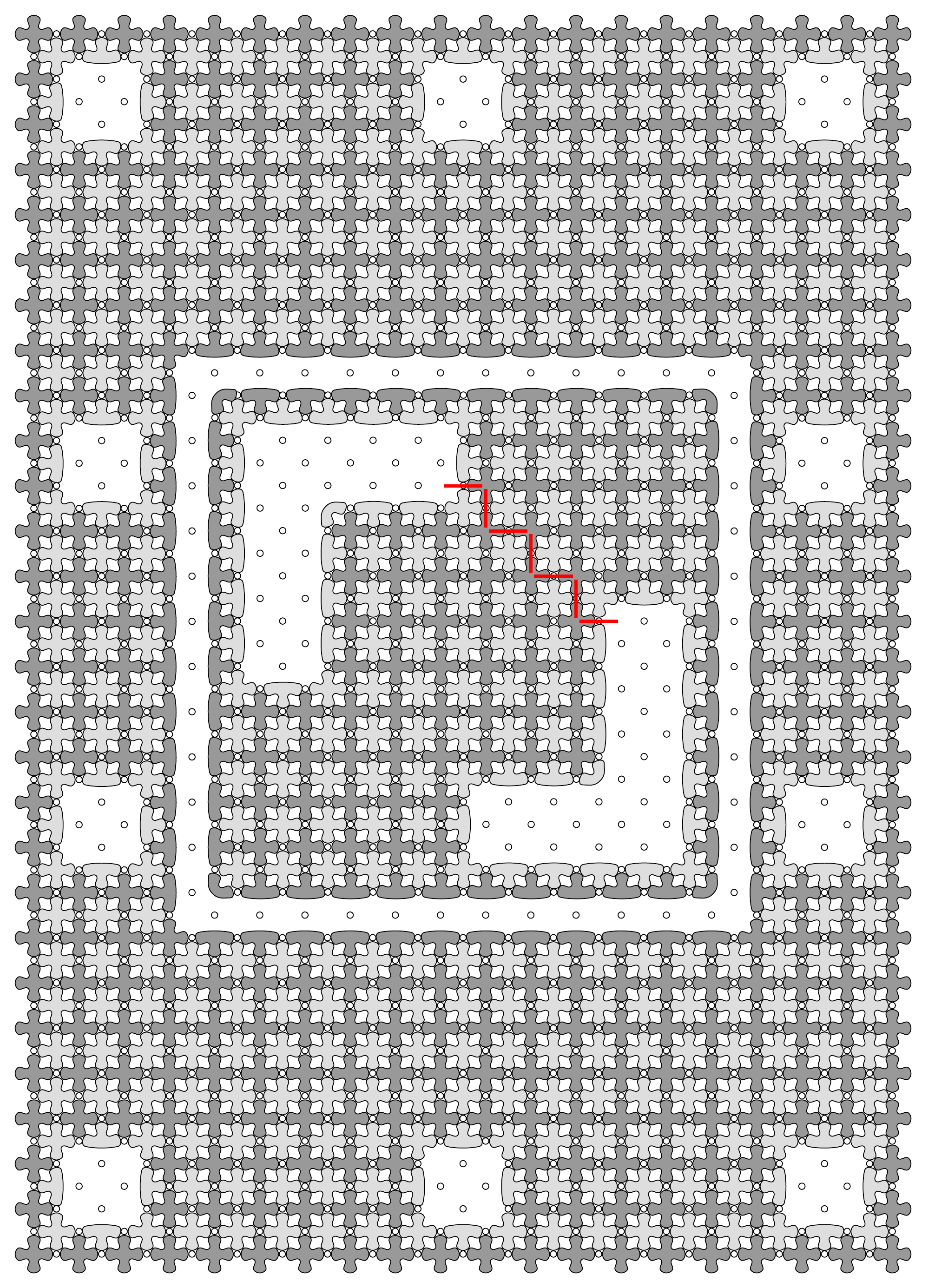}}
\end{center}
\caption{The defects are expanded, beginning the process of moving them back to their original position. They cannot be moved further than as shown without reducing the distance of the code. A minimum length error chain is shown. Note that there are no errors that lead to diagonal segments in this direction \cite{Wang11}.}\label{Hadamard_G}
\end{figure*}

\begin{figure*}
\begin{center}
\resizebox{150mm}{!}{\includegraphics{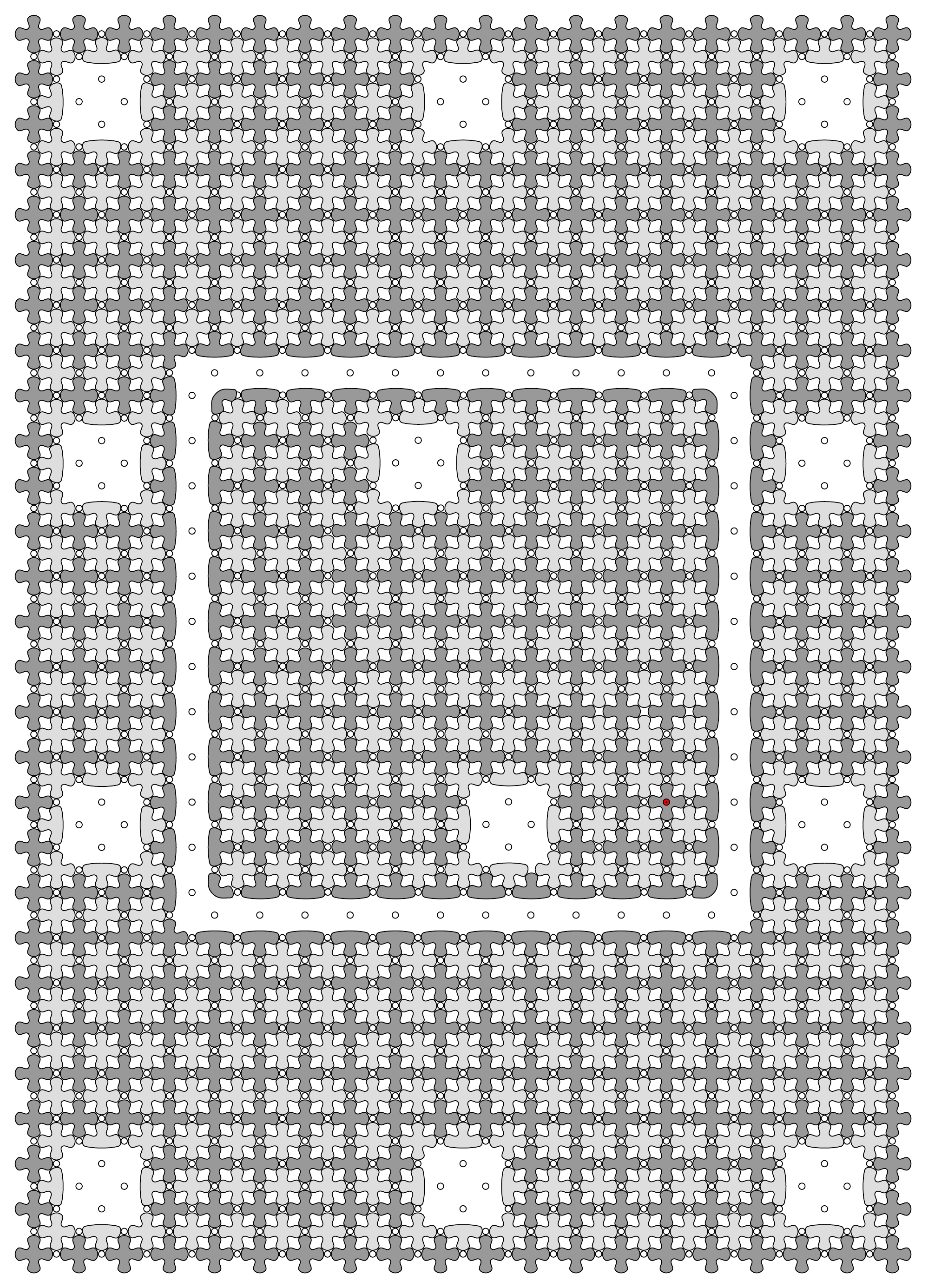}}
\end{center}
\caption{After a second round of stabilizer measurement in the expanded defect pattern, a number of stabilizer checks are turned back on to reduce the size of these defects and partially complete movement of these defects back to their original position. The first segment of a temporal error chain is shown.}\label{Hadamard_H}
\end{figure*}

\begin{figure*}
\begin{center}
\resizebox{150mm}{!}{\includegraphics{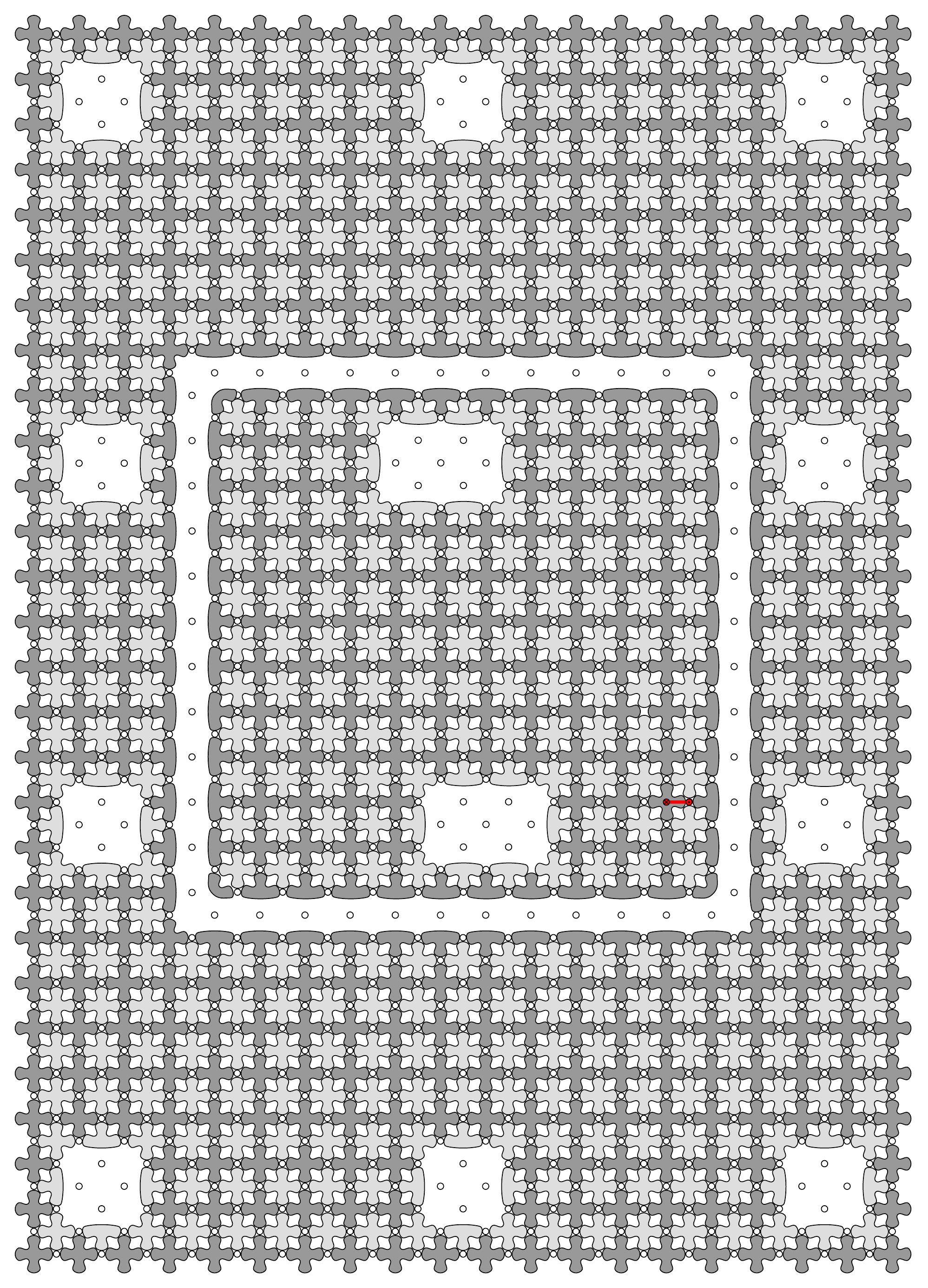}}
\end{center}
\caption{The defects are expanded to include their original locations. This pattern is applied twice. Half a diagonal space-time error segment is shown and a full temporal segment located at the left end is implied.}\label{Hadamard_I}
\end{figure*}

\begin{figure*}
\begin{center}
\resizebox{150mm}{!}{\includegraphics{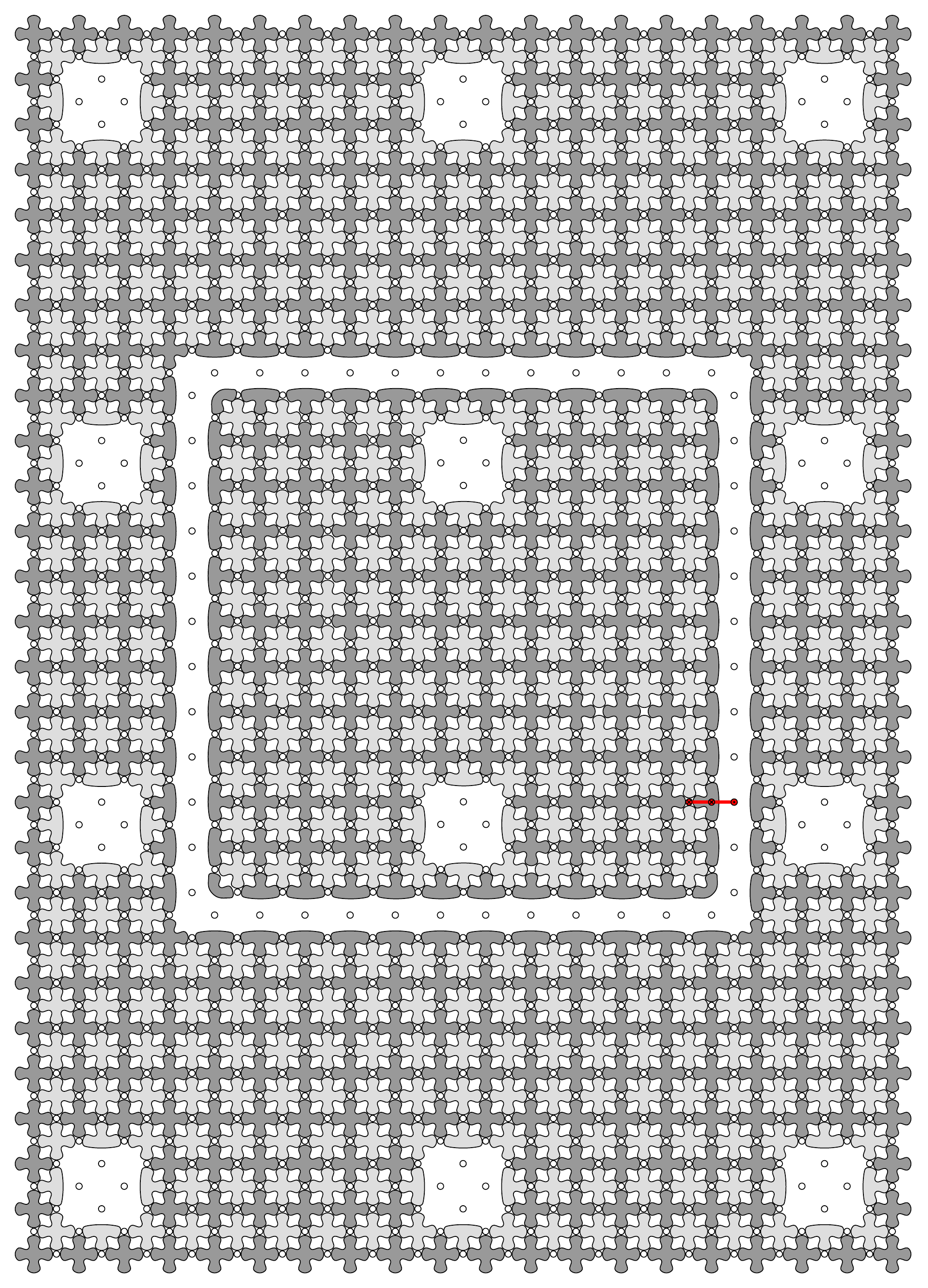}}
\end{center}
\caption{The defects are contracted back to their original sizes in their original locations. This measurement pattern is applied three times. Two half space-time error segments are shown and an additional two centrally located temporal segment implied.}\label{Hadamard_J}
\end{figure*}

\begin{figure*}
\begin{center}
\resizebox{150mm}{!}{\includegraphics{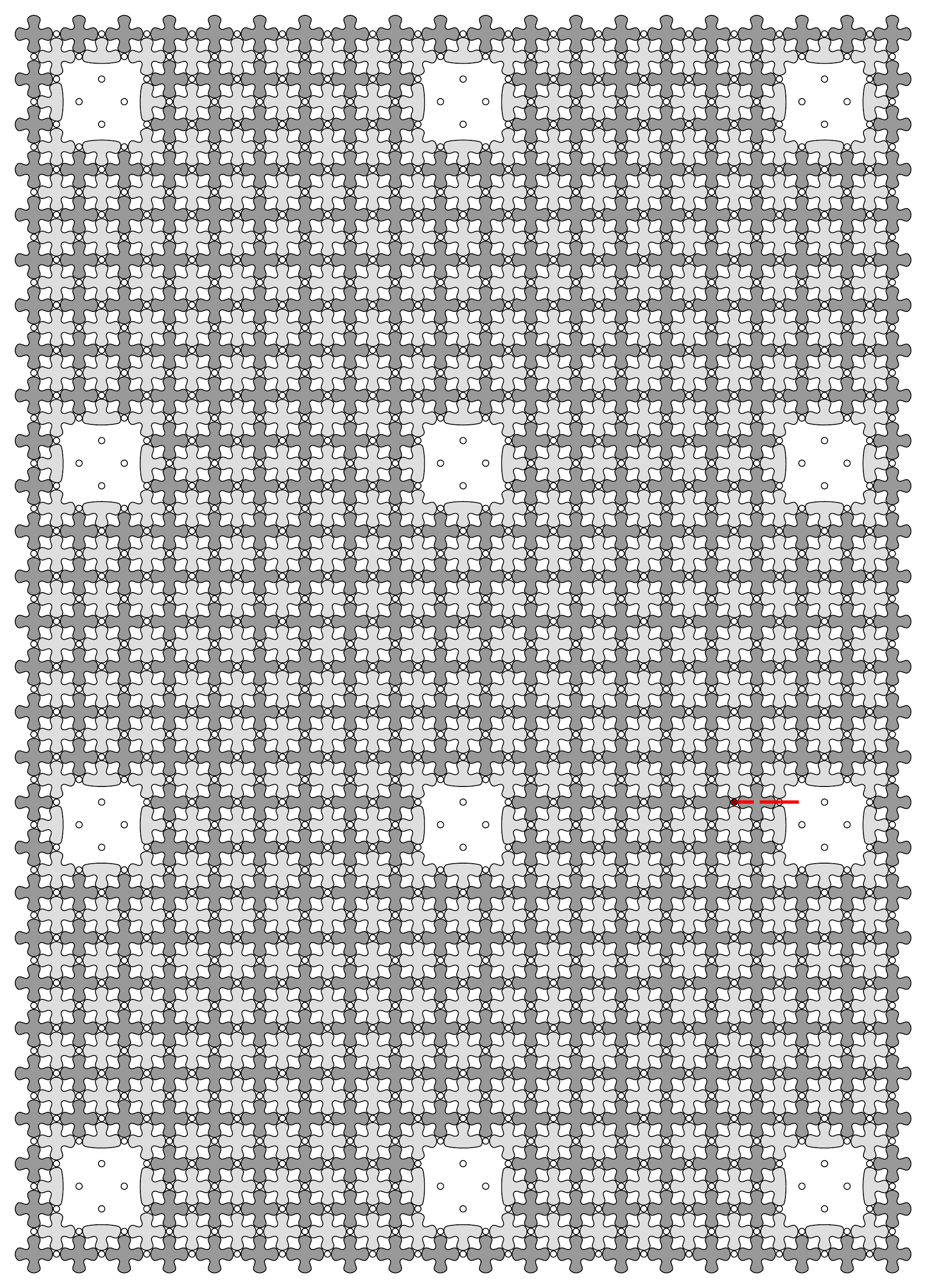}}
\end{center}
\caption{All remaining stabilizers are restored to their original weight reintegrating the logical qubit that has undergone logical $H$ with the rest of the computer. It can be directly verified that the end of the error chain shown consists of a total of 7 segments, implying the distance of the code has been maintained.}\label{Hadamard_K}
\end{figure*}

Previous implementations of surface code logical $H$ required braiding and significant space around the logical qubit \cite{Fowl08}. Our new implementation is described in Figs.~\ref{Hadamard_A}--\ref{Hadamard_K} and does not suffer either of these drawbacks.

\section{Acknowledgements}

This research was conducted by the Australian Research Council Centre of Excellence for Quantum Computation and Communication Technology (project number CE110001027), with support from the US National
Security Agency and the US Army Research Office under
contract number W911NF-08-1-0527. Supported by the Intelligence Advanced Research Projects Activity (IARPA) via Department of Interior National Business Center contract number D11PC20166.  The U.S. Government is authorized to reproduce and distribute reprints for Governmental purposes notwithstanding any copyright annotation thereon.  Disclaimer: The views and conclusions contained herein are those of the authors and should not be interpreted as necessarily representing the official policies or endorsements, either expressed or implied, of IARPA, DoI/NBC, or the U.S. Government.

\bibliography{../References}

\end{document}